\documentclass[fleqn,usenatbib,useAMS]{mnras}
\usepackage{graphicx}	
\usepackage{amsmath}
\usepackage{amssymb}
\usepackage{multicol}
\usepackage{bm}		
\usepackage{pdflscape}	
\usepackage{orcidlink}
\usepackage[T1]{fontenc}
\usepackage{ae,aecompl}
\usepackage{newtxtext,newtxmath}
\graphicspath{{./}{figures/}}
\title[Accretion state in HFQPOs]{Revisiting the observation-theory confrontation in high-frequency QPOs: a new QPO in NGC 5506, intermediate mass black holes, and the crucial role of accretion state}
\author[Zhang et al.]{Haoyang Zhang\orcidlink{0000-0003-3392-320X}$^{1,2}$, Lingwei Meng$^{1}$, Li Zhang$^{1,2}$ and Benzhong Dai\orcidlink{0000-0001-7908-4996}$^{1,2}$\thanks{Corresponding author: \href{mailto:bzhdai@ynu.edu.cn}{bzhdai@ynu.edu.cn}}
	\\
	$^{1}$Department of Astronomy, Yunnan University, Kunming 650091, China\\
	$^{2}$Key Laboratory of Astroparticle Physics of Yunnan Province, Yunnan University, Kunming 650091, China
}
\date{Accepted XXX. Received YYY; in original form ZZZ}
\pubyear{2025}
\begin{document}
\maketitle
\begin{abstract}
The scale invariance of accretion processes (SIAP) is crucial for understanding the physical processes of black hole accretion systems at different scales. When applying this rule to high-frequency quasi-periodic oscillations (HFQPOs), there is an observation--theory confrontation in active galactic nuclei (AGNs). By compiling an updated X-ray HFQPO catalog, we found that the ultraluminous X-ray sources support the HFQPO models, similar to black hole X-ray binaries. More importantly, we identified two supermassive black hole (SMBH) sources (Sgr A* and NGC 1365) with possible advection-dominated accretion flow (ADAF) configurations that support existing HFQPO models, even though many AGNs still do not. Furthermore, we report a new HFQPO candidate in NGC 5506. This source exhibits an accretion state similar to that of Sgr A* and NGC 1365, and it also supports the HFQPO models. Our results are consistent with previous numerical simulations and suggest that the accretion state of HFQPOs in SMBHs may differ from that of stellar-mass black holes (SBHs). To reconcile the sources that do not support the models, either a global general-relativistic HFQPO model based on magnetohydrodynamics needs to be considered, or the HFQPOs in these sources may originate from entirely different physical processes. This discovery significantly extends the SIAP rule to a broader scale, confirming that the paradigm of accretion scale invariance remains consistent from SBHs to SMBHs.
\end{abstract}

\begin{keywords}
Galaxies: Seyfert --- X-rays: galaxies --- X-rays: binaries --- black holes physics---accretion, accretion disks
\end{keywords}

\section{Introduction} \label{sec:intro}
Active galactic nuclei (AGNs; \citealt{1995PASP..107..803U}) are thought to have a supermassive black hole (SMBH; $10^{6-10}M_{\sun}$) accretion system in their center. Some of them exhibit jet structures. Black hole X-ray binaries (XRBs) containing a stellar-mass black hole (SBH; 5--15$M_{\sun}$) can accrete matter, forming a micro-accretion system from its companion star \citep{Tauris}. In addition to the physical structure, both AGNs and XRBs have many observational similarities, such as bright X-ray emissions \citep{2006ARA&A..44...49R,2017A&ARv..25....2P}, radio jet structures \citep{1995PASP..107..803U,1999ARA&A..37..409M}, radio--X-ray correlations \citep{2006A&A...456..439K}, and X-ray quasi-periodic oscillations (QPOs; \citealt{2006ARA&A..44...49R,2008Natur.455..369G}). The ultraluminous X-ray sources (ULXs; \citealt{2003ApJ...585L..37M,2004ApJS..154..519S,2015ApJ...809L..14B,2017ARA&A..55..303K}) which potentially contain intermediate-mass black holes (IMBH; $10^{2-5}M_{\sun}$), have also exhibited the QPO phenomenon \citep{2007ApJ...660..580S,2014Natur.513...74P,2015ApJ...811L..11P}. However, it has also been argued that some of them contain only a stellar-mass compact object \citep{2014Natur.514..202B,2017MNRAS.466L..48I,2020ApJ...895...60R}. It seems that X-ray QPO is a common observational property in these accretion systems with different central black hole masses. 

As the sample size increased, a scaling relationship among X-ray luminosity, radio luminosity, and black hole mass was discovered and extended from SBHs to SMBHs, known as the fundamental plane of black hole activity \citep{2003MNRAS.345.1057M,2006A&A...456..439K,2016MNRAS.455.2551N,Romero_2017}. Furthermore, the discovery of mass-scaled characteristic damping timescales of the accretion disk and the jet \citep{2006Natur.444..730M,2021Sci...373..789B,2024ApJ...967L..18Z} further indicates that these objects differ only in black hole scales. Additionally, the radiative properties of accretion disks and jets in SBHs and SMBHs exhibit remarkable similarities \citep{Romero_2017,2021MNRAS.500.2475J}. These findings collectively suggest that black holes across different mass scales follow scaling relationships and share analogous accretion properties, a phenomenon referred to as the scale invariance of accretion processes (SIAP).

In the 1970s, \citet{1973A&A....24..337S} introduced the standard accretion disk model, commonly referred to as the Shakura--Sunyaev disk (SSD), which describes an accretion process where angular momentum is transferred outward due to viscosity. This model, characterized by high radiative cooling efficiency, successfully explains the bright thermal radiation observed in some AGNs and the soft and  steep power law (SPL) states of XRBs \citep{1981ARA&A..19..137P,2006ARA&A..44...49R,2008bhad.book.....K}. For low-luminosity AGNs and hard state XRBs, the accretion flow in the inner region is typically described by advection-dominated accretion flows (ADAFs), which are radiatively inefficient with $\lambda=L_{\text{bol}}/L_{\text{Edd}}\lesssim10^{-2}$ \citep{1998tbha.conf..148N,2008ARA&A..46..475H,2014ARA&A..52..529Y}. As the disk thickness and internal temperature increase, material becomes ionized, producing non-thermal X-ray radiation through the Comptonization process. In such scenarios, the inward advection of material enhances the magnetic field near the black hole, potentially leading to the formation of a magnetically arrested disk \citep{2003ApJ...592.1042I,2003PASJ...55L..69N,2023Sci...381..961Y}. Strong magnetic fields facilitate jet generation by enabling energy extraction processes described in the Blandford--Znajek \citep{1977MNRAS.179..433B} and Blandford--Payne \citep{1982MNRAS.199..883B} processes. Furthermore, numerical simulations indicate distinct accretion states between SSDs and ADAFs, highlighting the complex interplay between accretion processes, radiative efficiency, and jet formation \citep{2014ARA&A..52..529Y}.

The X-ray QPO phenomenon initially detected in XRBs can be divided into high-frequency QPOs (HFQPOs; 40--450 Hz) and low-frequency QPOs (LFQPOs; 0.1--30 Hz). In general, HFQPO is associated with the SPL state \citep{2006ARA&A..44...49R}. HFQPOs in some sources exhibited transient non-simultaneous 3:2 harmonics (e.g., \citealt{2004AIPC..714...13R,1999ApJ...522..397R,2002ApJ...580.1030R}). Further studies have detected HFQPOs with high significance in several narrow-line Seyfert 1 galaxies (a subclass of AGNs with high accretion rates ($\lambda\gtrsim10^{-1}$), \citealt{komossa2007narrowlineseyfert1galaxies}), with periods ranging from one to a few hours (e.g., \citealt{2008Natur.455..369G,2015MNRAS.449..467A,2017ApJ...849....9Z,2018AA...616L...6G,2018ApJ...853..193Z,2020MNRAS.495.3538J,2020AcASn..61....2Z,2024AA...691A...7Y}; see also Table 2 in \citealt{2021ApJ...906...92S}). Harmonics with a 2:1 ratio have only been detected in 1H 0707-495 and ESO 113-G010 \citep{2018ApJ...853..193Z,2020AcASn..61....2Z}. Of these reports, only the QPO signal from RE J1034+396 is considered significant and persistent \citep{2012A&A...544A..80G,2014MNRAS.445L..16A,2020MNRAS.495.3538J,2023ApJ...946...52Z,2024ApJ...961L..32X}. In addition, \citet{2024ApJ...961L..32X} discovered an interesting relationship between time lag and QPO frequency in RE J1034+396. QPO phenomena were also observed in ULXs, ranging from approximately $10^{-2}$ to a few hertz. Harmonics with a 3:2 ratio were also detected in ULXs \citep{2014Natur.513...74P,2015ApJ...811L..11P} and Sgr A* \citep{2004AA...425.1075A}. LFQPOs are more stable and persistent than HFQPOs \citep{2001ApJ...556..515M}. It has been observed in both the hard and the SPL states of XRBs \citep{2006ARA&A..44...49R}. So far, LFQPOs in ULXs have not been reported, and only one has been reported in AGNs \citep{2018ApJ...860L..10S}.

\citet{2015ApJ...798L...5Z} suggested that the QPOs of ULXs and AGNs are analogous to HFQPOs in IMBH and SMBH scales by using the relationship ($f_{\text{QPO}}\propto M_{\text{BH}}$), known as the linear universal scaling relation. In what follows, we refer to the QPOs of ULXs and AGNs as HFQPOs.

The exact mechanism of HFQPO is currently uncertain. A popular view is the resonance model, a coupled state scenario with 3:2 ratio oscillation frequencies from various directions in accretion disk \citep{2001A&A...374L..19A,2003PASJ...55..467A,2004AA...425.1075A}. Further observations extended this applicability to the X-ray HFQPOs of ULXs and AGNs \citep{2015ApJ...798L...5Z} despite harmonics not being detected in most of these. Other physical explanations of HFQPO have been proposed and applied to XRBs and AGNs: instability at the disk--magnetosphere interface \citep{2004ApJ...601..414L,2012MNRAS.423.3083M}, the hotspot model \citep{1999ApJ...524L..63S,2004ApJ...606.1098S,2023ApJ...946...52Z}, the dynamical transient chaos model \citep{1993ApJ...411L..91S}, diskoseismology \citep{1997ApJ...476..589P,1999PhR...311..259W,2013IAUS..290...57L}, and the magnetic reconnection of the inner and outer disc regions \citep{2009MNRAS.398.1886Z}. LFQPOs could be induced by the Lense--Thirring precession of an ADAF \citep{1998ApJ...492L..59S,2019NewAR..8501524I}. Alternative models have also been proposed \citep{2000ApJ...531L..41C,2000ApJ...538L.137N,2000ApJ...542L.111T}.

Above physical mechanisms are based on the accretion disk and central black hole. Therefore, the successful application of HFQPO models to ULX and AGN can also indicate the SIAP. Those scenarios can even be used to measure the dimensionless spin ($a=cJ/GM_{\text{BH}}^{2}$, hereafter ``spin'') and mass of black holes \citep{2015ApJ...798L...5Z}. In addition, all three types of objects exhibit similar states of high accretion rates observationally, providing further support for the SIAP framework \citep{2006ARA&A..44...49R,komossa2007narrowlineseyfert1galaxies,2014Natur.513...74P}. However, when the spin of the black hole (obtained as an independent measurement by other methods) is included in the comparison of observations and theoretical models, a confrontation arises \citep{2021ApJ...906...92S}. Specifically, XRBs fit well with the above physical scenarios, although AGNs are unable to follow them. It is worth noting that ULXs and Sgr A* were not included in their study. Thus, either HFQPO models need to incorporate additional physical parameters (as SMBHs differ from SBHs in aspects such as space-time environment and higher likelihood of chaotic accretion), or the physical origin of HFQPOs in AGNs is entirely different from that in XRBs. \citet{2024MNRAS.532.2143K} suggested that a binary system containing SBH and SMBH, also known as extreme mass ratio inspirals (EMRIs), can produce X-ray HFQPOs in AGNs, which were applied to RE J1034+396.

Early theoretical studies focused on p-mode and g-mode, and successfully explained the QPO signal under an SSD configuration by considering the tilt and warping of the accretion disk \citep{2004PASJ...56..905K,2008MNRAS.386.2297F,2009ApJ...706..705H}. Further numerical simulations have been critical of these schemes \citep{2006ApJ...645L..65A,2009ApJ...692..869R}. Notably, to produce HFQPOs for Sgr A*, the SMBH counterpart at the center of Milky Way, these simulations require an ADAF disk configuration characterized by hot accretion flows \citep{2012ApJ...746L..10D,2012MNRAS.423.3083M,2013ApJ...774L..22S}. This scenario stands in stark contrast to the accretion states typically associated with most HFQPO sources, yet it provides a plausible explanation for the low accretion rate of HFQPOs observed in Sgr A* ($\lambda\sim10^{-8}$, \citealt{2009ApJ...697...45B}). Consequently, accretion states appear to play a critical role in the HFQPOs of SMBHs, although the specific mechanisms remain poorly understood. Direct observational evidence may be the key to unraveling this.

In this work, we compile an updated X-ray HFQPO catalog and revisit the confrontation between observations and theoretical models. In Section \ref{sec:obs_the}, we compare the updated catalog with existing HFQPO models and identify two sources, Sgr A* and NGC 1365, that support these models. In Section \ref{sec:qpo_cand}, we report a new X-ray HFQPO candidate, NGC 5506, which may exhibit an ADAF configuration similar to Sgr A* and NGC 1365. In Section \ref{sec:dis}, we discuss the role of accretion states in the presence of HFQPOs in SMBH systems and argue that the SIAP framework should consistently hold, as it is supported by HFQPO observations. Section \ref{sec:con} concludes.

\begin{table*}
	\centering
	\caption{Updated X-ray HFQPO catalog.}
	\begin{tabular}{ccccccl}
		\hline
		Name & Type & Log $M_{\text{BH}}$ & Dimensionless spin of black hole & $f_{\text{QPO}}$ & Harmonic signal & Ref of $M_{\text{BH}}$--$a$--$f_{\text{QPO}}$ \\ & & $M_{\sun}$ & $a$ & Hz & & \\[1ex]
		\hline
		XTE J1550-564 & XRB & $0.96^{+0.03}_{-0.03}$  & 0.29--0.62, 0.75--0.77                          & 184                    & Yes & 1 -- 20, 21 -- 40 \\[1ex]
		XTE J1650-500 & XRB & $0.70^{+0.15}_{-0.22}$  & 0.78--0.80                                      & 250                    & No  & 2 -- 21 -- 41 \\[1ex]
		XTE J1859+226 & XRB & $0.89^{+0.10}_{-0.12}$  & 0.149$\pm{0.005}$,0.982--0.990                  & 190                    & No  & 3 -- 22, 23 -- 42 \\[1ex]
		4U 1630–47    & XRB & 1.00$\pm{0.004}$        & $0.985^{+0.005}_{-0.014}$, 0.92$\pm{0.04}$      & 184                    & No  & 4 -- 24, 25 -- 43 \\[1ex]
		GRO J1655-40  & XRB & $0.73^{+0.03}_{-0.02}$  & 0.65--0.75, 0.90--0.998, 0.97--0.99             & 300                    & Yes & 5 -- 26, 27, 21 -- 40, 44\\[1ex]
		GRS 1915+105  & XRB & $1.10^{+0.06}_{-0.07}$  & 0.97--0.99                                      & 41, 113                & Yes & 6 -- 28, 29, 30 -- 43 \\[1ex]
		H1743-322     & XRB & $1.04^{+0.07}_{-0.07}$  & $0.67^{+0.23}_{-0.15}$                          & 163                    & Yes & 7 -- 31 -- 45 \\[1ex]
		M82 X-1       & ULX & $2.62^{+0.04}_{-0.07}$  & 0.18$\pm{0.12}$                                 & 3.32                   & Yes & 8 -- 8 -- 8 \\[1ex]
		NGC 5408 X-1  & ULX & $5.0^{+0.11}_{-0.15}$   & 0.998, 0.95                                     & 2.0 $\times\ 10^{-2}$  & No  & 9 -- 9 -- 46 \\[1ex]
		Sgr A*        &GCBH & $6.63^{+0.04}_{-0.04}$  & 0.90$\pm{0.06}$                                 & 1.07 $\times\ 10^{-3}$ & Yes & 10 -- 32 -- 47 \\[1ex]
		ASASSN-14li   & TDE & $6.23^{+0.35}_{-0.35}$  & \textgreater\ 0.7                               & 7.7 $\times\ 10^{-3}$  & No  & 11 -- 33 -- 33 \\[1ex]
		Ton S 180     & AGN & $6.85^{+0.50}_{-0.50}$  & \textless\ 0.4                                  & 5.56 $\times\ 10^{-6}$ & No  & 12 -- 34 -- 48 \\[1ex]
		ESO 113-G010  & AGN & $6.85^{+0.15}_{-0.24}$  & 0.996                                           & 1.24 $\times\ 10^{-4}$ & Yes & 13 -- 13 -- 49 \\[1ex]
		1H0419-577    & AGN & $8.11^{+0.50}_{-0.50}$  & \textgreater\ 0.98                              & 2.0 $\times\ 10^{-6}$  & No  & 14 -- 35 -- 48 \\[1ex]
		1H0707-495    & AGN & $6.36^{+0.24}_{-0.06}$  & \textgreater\ 0.976                             & 2.6 $\times\ 10^{-4}$  & Yes & 15 -- 36 -- 50 \\[1ex]
		RE J1034+396  & AGN & $6.5^{+0.50}_{-0.50}$   & 0.998                                           & 2.7 $\times\ 10^{-4}$  & No  & 16 -- 16 -- 51 \\[1ex]
		Mrk 766       & AGN & $6.82^{+0.05}_{-0.06}$  & \textgreater\ 0.92                              & 1.55 $\times\ 10^{-4}$ & No  & 17 -- 37 -- 52 \\[1ex]
		MCG-06-30-15  & AGN & $6.20^{+0.09}_{-0.12}$  & \textgreater\ 0.917                             & 2.73 $\times\ 10^{-4}$ & No  & 18 -- 38 -- 53 \\[1ex]
		NGC 1365      & AGN & $6.87^{+0.13}_{-0.17}$  & $0.97^{+0.01}_{-0.04}$                          & 2.19 $\times\ 10^{-4}$ & No  & 19 -- 39 -- 54 \\[1ex]
		\hline
	\end{tabular}
	{Notes---References:
		(1) \citet{2011ApJ...730...75O}, (2) \citet{2004ApJ...616..376O}, (3) \citet{2022MNRAS.517.1476Y}, (4) \citet{2014ApJ...789...57S}, (5) \citet{2002MNRAS.331..351B}, (6) \citet{2014ApJ...796....2R}, (7) \citet{2016cosp...41E1324M}, (8) \citet{2014Natur.513...74P}, (9) \citet{2013RAA....13..705H}, (10) \citet{2009ApJ...692.1075G}, (11) \citet{2016Sci...351...62V}, (12) \citet{2012ApJ...754..146M}, (13) \citet{2013ApJ...764L...9C}, (14) \citet{2010ApJS..187...64G}, (15) \citet{2005ApJ...618L..83Z}, (16) \citet{2016AA...594A.102C}, (17) \citet{2015PASP..127...67B}, (18) \citet{2016ApJ...830..136B}, (19) \citet{2019AA...622A.128F}, (20) \citet{2011MNRAS.416..941S}, (21) \citet{2009ApJ...697..900M}, (22) \citet{2022MNRAS.517.1469M}, (23) \citet{2024MNRAS.52712053M}, (24) \citet{2014ApJ...784L...2K}, (25) \citet{2018ApJ...867...86P}, (26) \citet{2006ApJ...636L.113S}, (27) \citet{2009MNRAS.395.1257R}, (28) \citet{2009ApJ...706...60B}, (29) \citet{2013ApJ...775L..45M}, (30) \citet{2006ApJ...652..518M}, (31) \citet{2022Univ....8..273D}, (32) \citet{2024MNRAS.527..428D}, (33) \citet{2019Sci...363..531P}, (34) \citet{2018MNRAS.474.1538P}, (35) \citet{2019MNRAS.483.2958J}, (36) \citet{2010MNRAS.401.2419Z}, (37) \citet{2018MNRAS.480.3689B}, (38) \citet{2007PASJ...59S.315M}, (39) \citet{2013Natur.494..449R}, (40) \citet{2002ApJ...580.1030R}, (41) \citet{2003ApJ...586.1262H}, (42) \citet{2000ApJ...535L.123C}, (43) \citet{2004AIPC..714...13R}, (44) \citet{1999ApJ...522..397R}, (45) \citet{2005ApJ...623..383H}, (46) \citet{2007ApJ...660..580S}, (47) \citet{2004AA...425.1075A}, (48) \citet{2003ApJ...585..665H}, (49) \citet{2020AcASn..61....2Z}, (50) \citet{2018ApJ...853..193Z}, (51) \citet{2008Natur.455..369G}, (52) \citet{2017ApJ...849....9Z}, (53) \citet{2018AA...616L...6G}, (54) \citet{2024AA...691A...7Y}.}
	\label{tab:catalog}
\end{table*}

\begin{figure*}
	\centering
	\includegraphics[width=\linewidth]{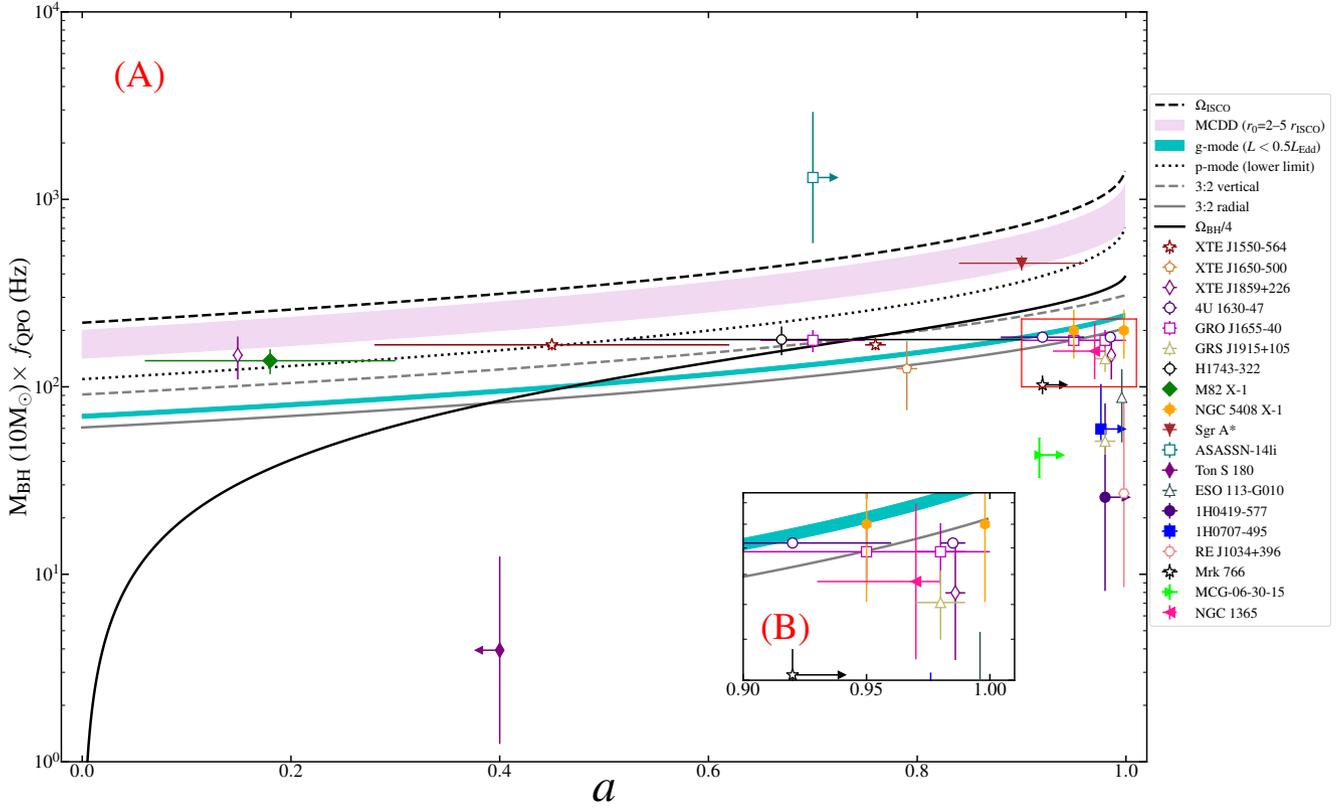}
	\caption{
		Comparison between models and observations with spin. Panel (B) is the magnified view of the region within the red box in panel (A). The black dashed line, light pink region, gray dotted line, gray dashed line, gray solid line, black solid line, and skyblue bold line denote the orbital frequency in $r_{\text{ISCO}}$, MCDD model of $r_{0}$=2--5$\ r_{\text{ISCO}}$, p-mode oscillation, vertical frequency of the 3:2 resonance model, radial frequency of the 3:2 resonance model, magnetospheric interface oscillation, and g-mode oscillation, respectively.
	}
	\label{fig:model_fixed}
\end{figure*}

\section{Revisiting the observation--theory confrontation}\label{sec:obs_the}
\subsection{An updated X-ray HFQPO catalog}
Here, we further explore the SIAP and physical origins of HFQPOs. Based on \citet{2015ApJ...798L...5Z} and \citet{2021ApJ...906...92S}, we compiled an updated X-ray HFQPO catalog---see Table \ref{tab:catalog}---by adding some recent work (including NGC 5506). Several sources without mass and spin estimates are not shown in the table. Those sources can be found in \citet{2015ApJ...798L...5Z,2015ApJ...811L..11P,2021ApJ...906...92S}. This catalog contains almost all black hole counterparts of different scales, such as XRBs, ULXs, and AGNs. The Galactic Center black hole (GCBH) Sgr A* also shows a 3:2 harmonic signal \citep{2004AA...425.1075A,2005A&A...440....1T}. In keeping with previous studies, we only display two times the fundamental frequencies of these harmonic signals in Table \ref{tab:catalog}. It is important to note that the QPOs of Ton S 180 and 1H0419-577 have lower frequencies, raising uncertainty about their classification as HFQPOs. Since these sources are already included in \citet{2021ApJ...906...92S} and their accretion disks also emit extreme ultraviolet (EUV) radiation, we have included Ton S 180 and 1H0419-577 as EUV QPO sources in our catalog. For comparison, our catalog also includes the well-known QPO source from the TDE event, ASASSN-14li.
\subsection{Physical Mechanisms}
Here, we invoke several popular models in comparison. The hotspot model represents as a hotspot the Kepler motion in the innermost stable circular orbit (ISCO) of an accretion disk \citep{2004ApJ...606.1098S}. The formula of the Kepler motion is
\begin{equation}
	\Omega_{\text{K}}= (r^{3/2}+a)^{-1}\frac{c^{3}}{GM_{\text{BH}}},
\end{equation}
where $r$ is the radial position, $a$ is the spin of the black hole, $M_{\text{BH}}$ is the central back hole mass. In the resonance model, the 3:2 ratio oscillation can be excited by vertical and radial resonances of particles in an exact radial position \citep{2004AA...425.1075A}. The angular frequencies of two directions are given by
\begin{equation}
	\Omega_{\text{V}}= [\Omega_{\text{K}}^{2}(1-\frac{4a}{r^{3/2}}+\frac{3a^{2}}{r^{2}})]^{1/2},
\end{equation}
\begin{equation}
	\Omega_{\text{R}}= [\Omega_{\text{K}}^{2}(1-\frac{6}{r}+\frac{8a}{r^{3/2}}-\frac{3a^{2}}{r^{2}})]^{1/2}.
\end{equation}
The magnetic field at $r_{\text{ISCO}}$ can couple with magnetic fields at other radii, a phenomenon referred to as the magnetic connection of the inner and outer disk regions (MCDD). The differential rotation between these radii drives continual magnetic reconnection oscillations \citep{2009MNRAS.398.1886Z}. The oscillation frequency corresponds to the difference in Keplerian frequencies between the two radii, given by
\begin{equation}
	\Omega_{\text{MCDD}}= \Omega_{\text{ISCO}}-\Omega_{r_{0}},
\end{equation}
where $r_{0}$ denotes the outer radius that is magnetically coupled to $r_{\text{ISCO}}$. The diskoseismology can be divided into two modes. Gravity (g) mode can be characterized by the net gravitational-centrifugal force contributing to the restoring force. Considering the axisymmetric condition, oscillation comes from the position with the maximum of the radial frequency \citep{1997ApJ...476..589P}. The analytical solution is written as
\begin{equation}
	f_{\text{QPO}}= 714(M_{\sun}/M_{\text{BH}})F(a)[1-0.1\lambda], \label{equ:eq4}
\end{equation}
where $M_{\sun}$ is the mass of the Sun, $F(a)$ ranges in 1--3.443 for $a$=0--0.998 and $\lambda$ is the Eddington luminosity ratios. In pressure (p) mode, the restoring force can be attributed to pressure gradients. \citet{2013IAUS..290...57L} discussed this model in detail, with the lower limit expressed as $0.5\times\Omega_{\text{ISCO}}$. The last model we applied was proposed by \citet{2004ApJ...601..414L} and \citet{2012MNRAS.423.3083M}, the so-called magnetosphere interface instability model. The periodic signal could originate from Rayleigh--Taylor instability in the interface between the disk and central object. The angular frequency of QPO can be written as $\Omega_{\text{BH}}/4$, where $\Omega_{\text{BH}}= (c^{3}/GM_{\text{BH}})a/[2(1+\sqrt{1-a^{2}})]$ is the angular frequency of a rotating black hole \citep{2013ApJ...774L..22S}. We did not consider the dynamical transient chaos model because it is a phenomenological model without the black hole mass and spin as parameters \citep{1993ApJ...411L..91S}.

According to \citet{2021ApJ...906...92S}, we should add the spin of the black hole into the theoretical models. In the inner region of the accretion disk, the spin of the black hole has a strong effect on space-time \citep{1972ApJ...178..347B}. The $r_{\text{ISCO}}$ and spin have the following relationship:
\begin{equation}
\begin{aligned}
& r_{\text{ISCO}}=r_{\text{g}}[3+Z_{\text{2}}-\sqrt{(3-Z_{\text{1}})(3+Z_{\text{1}}+2Z_{\text{2}})}],\\
& Z_{\text{1}}\equiv 1+(1-a^{2})^{1/3}[(1+a^{2})^{1/3}+(1-a)^{1/3}],\\
& Z_{\text{2}}\equiv \sqrt{3a^{2}+Z_{\text{1}}^{2}},
\end{aligned}
\end{equation}
where $r_{\text{g}}=GM_{\text{BH}}/c^{2}$ is the gravitational radius.

In Figure \ref{fig:model_fixed}, only GRS 1915+105 is below the models, while other XRB and ULX sources are consistent with the theoretical models, despite different sources following different models. Most AGN data are located in the lower-right corner of the phase space and cannot support the models, similar to the results of \citet{2021ApJ...906...92S}.

Sgr A* is the first SMBH source with an ADAF configuration observed to follow the models in Figure \ref{fig:model_fixed}, consistent with results from numerical simulations \citep{2012ApJ...746L..10D,2013ApJ...774L..22S}. The Eddington luminosity ratios ($\lambda$) of AGNs exhibiting X-ray HFQPOs can be derived from the luminosity estimates of \citet{2012A&A...544A..80G} using the following equation:
\begin{equation}
	L_{\text{Edd}}=1.3\times10^{38}M_{\text{BH}}(M_{\sun})\ \text{erg/s}.
\end{equation}
The mean value of $\lambda$ is $\sim 0.43$, comparable to the typical value observed in narrow-line Seyfert 1 galaxies. By contrast, the $\lambda$ value of the recently reported HFQPO source \citep{2024AA...691A...7Y}, the Seyfert 1.8 AGN NGC 1365, is $0.065^{+0.031}_{-0.017}$, significantly lower than the average. Notably, panel (B) of Figure \ref{fig:model_fixed} shows that NGC 1365 occupies the same parameter space as ULXs and XRBs, distinctly separated from other AGNs. After accounting for the uncertainty in the y-axis value in Figure \ref{fig:model_fixed}, NGC 1365 approximately follows the HFQPO models. Additionally, \citet{1995A&A...295..585S} reported that this source may harbor a weak radio jet structure, further supporting an ADAF configuration with a low accretion rate. Therefore, both Sgr A* and NGC 1365 exhibit possible ADAF disk configurations and provide evidence supporting the HFQPO models. Two EUV QPO sources also fall outside the models. This may be because EUV radiation and X-rays originate at different locations in the accretion disk. Meanwhile, 1H0419-577 exhibits low radiation efficiency ($\lambda \sim 0.0075$), suggesting a possible ADAF configuration. This source is positioned in phase space above the super-Eddington accretion source Ton S 180 \citep{10.1093/mnras/staa2076} (see in Figure \ref{fig:model_fixed}) and similar to the findings for X-ray HFQPO sources. This suggests that the accretion state significantly effects QPOs, irrespective of the region of the accretion disk from which the signal originates. The QPO of TDE source occupies a distinct parameter space compared to other sources, favoring the hot spot model, which corresponds to the orbital period at the ISCO \citep{2019Sci...363..531P}.

\begin{figure*}
	\centering
	\includegraphics[width=\linewidth]{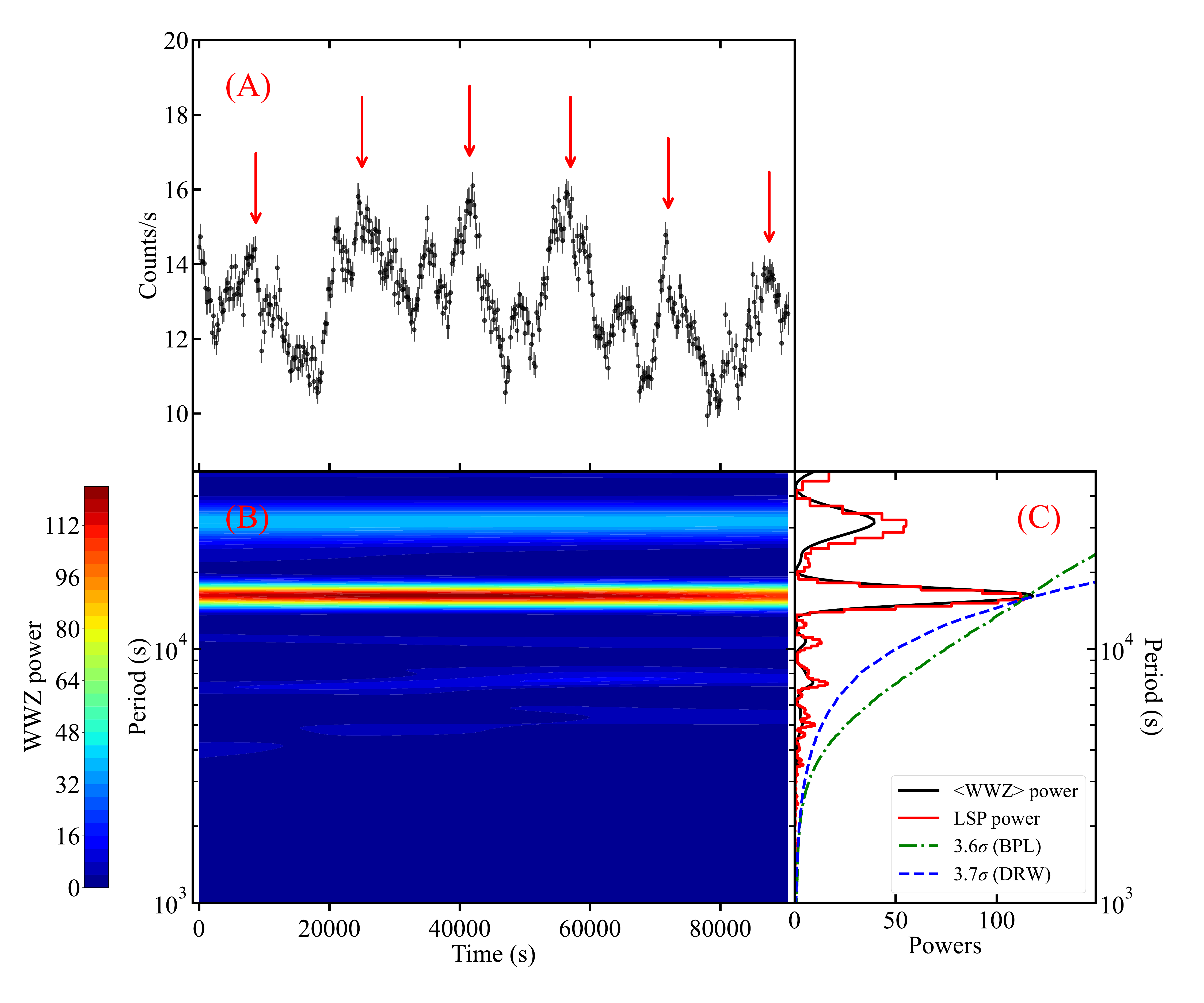}
	\caption{
		Results of WWZ and LSP analysis. Panel (A): light curve of ObsID:0554170201 in 0.2--10 keV with 200 s bin$^{-1}$. The red arrows indicate the peak positions of the periodic oscillation. A 2D time-frequency power map is shown in panel (B). The power spectra of LSP and time-averaged WWZ are respectively shown in red and black solid lines on panel (C).
	}
	\label{fig:wwz_result}
\end{figure*}

\begin{figure*}
	\includegraphics[width=\columnwidth]{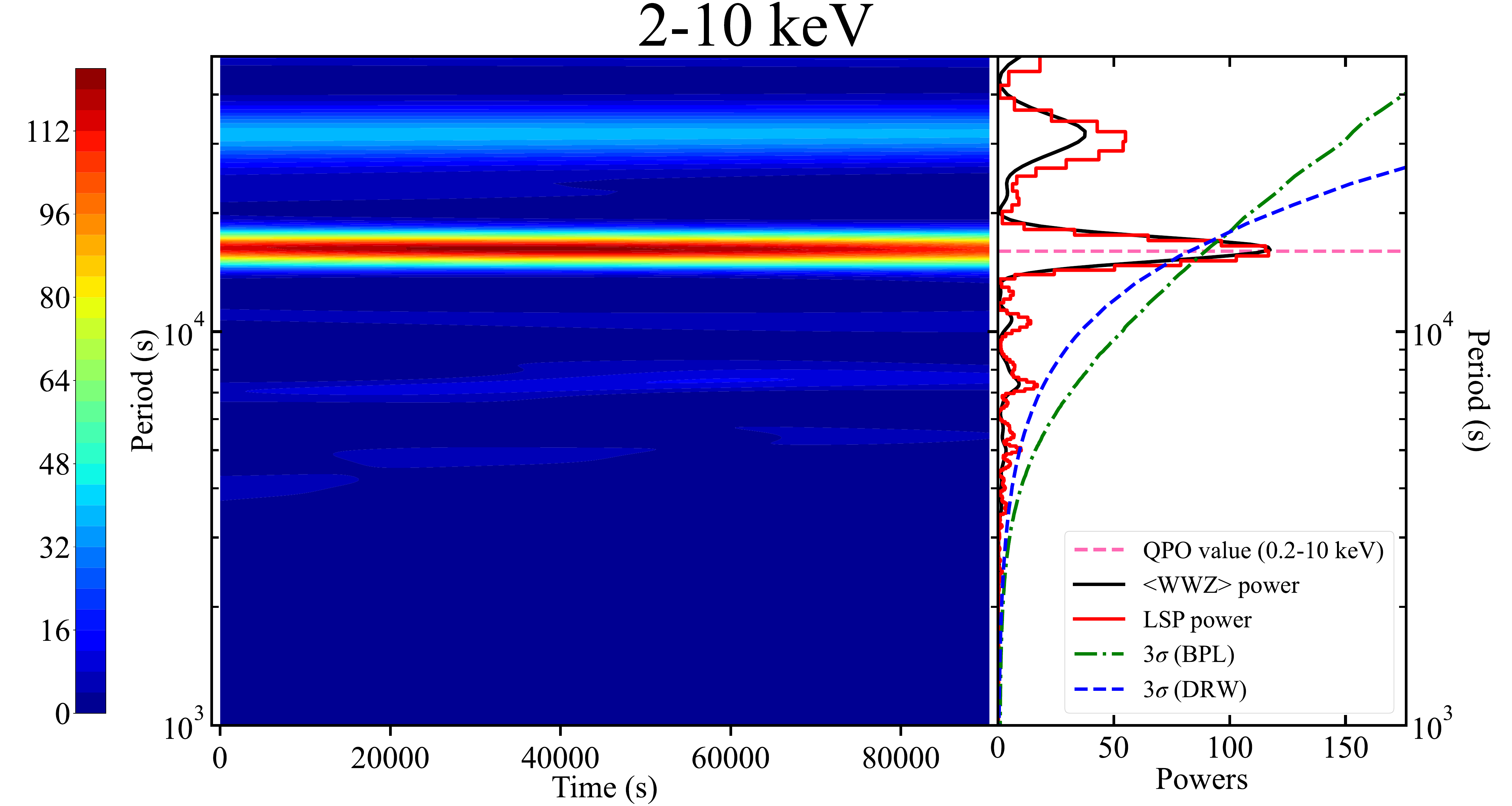}
	\includegraphics[width=\columnwidth]{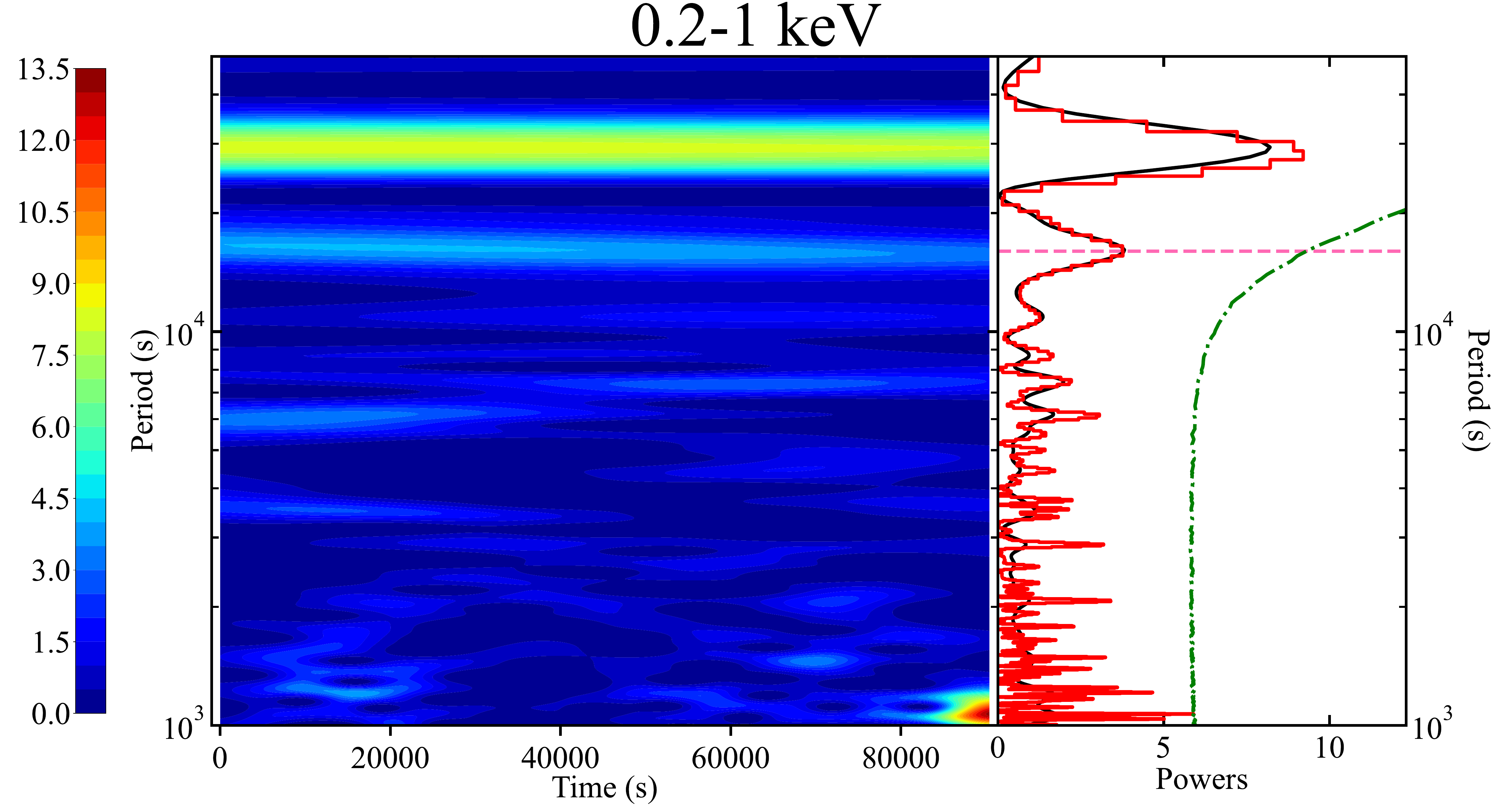}
	\caption{Results of WWZ and LSP analysis for the hard (left panel) and soft (right panel) bands of ObsID:0554170201. The pink dashed line represents the QPO value in the total band. \label{fig:wwz2}}
\end{figure*}

\begin{figure}
	\centering
	\includegraphics[width=\linewidth]{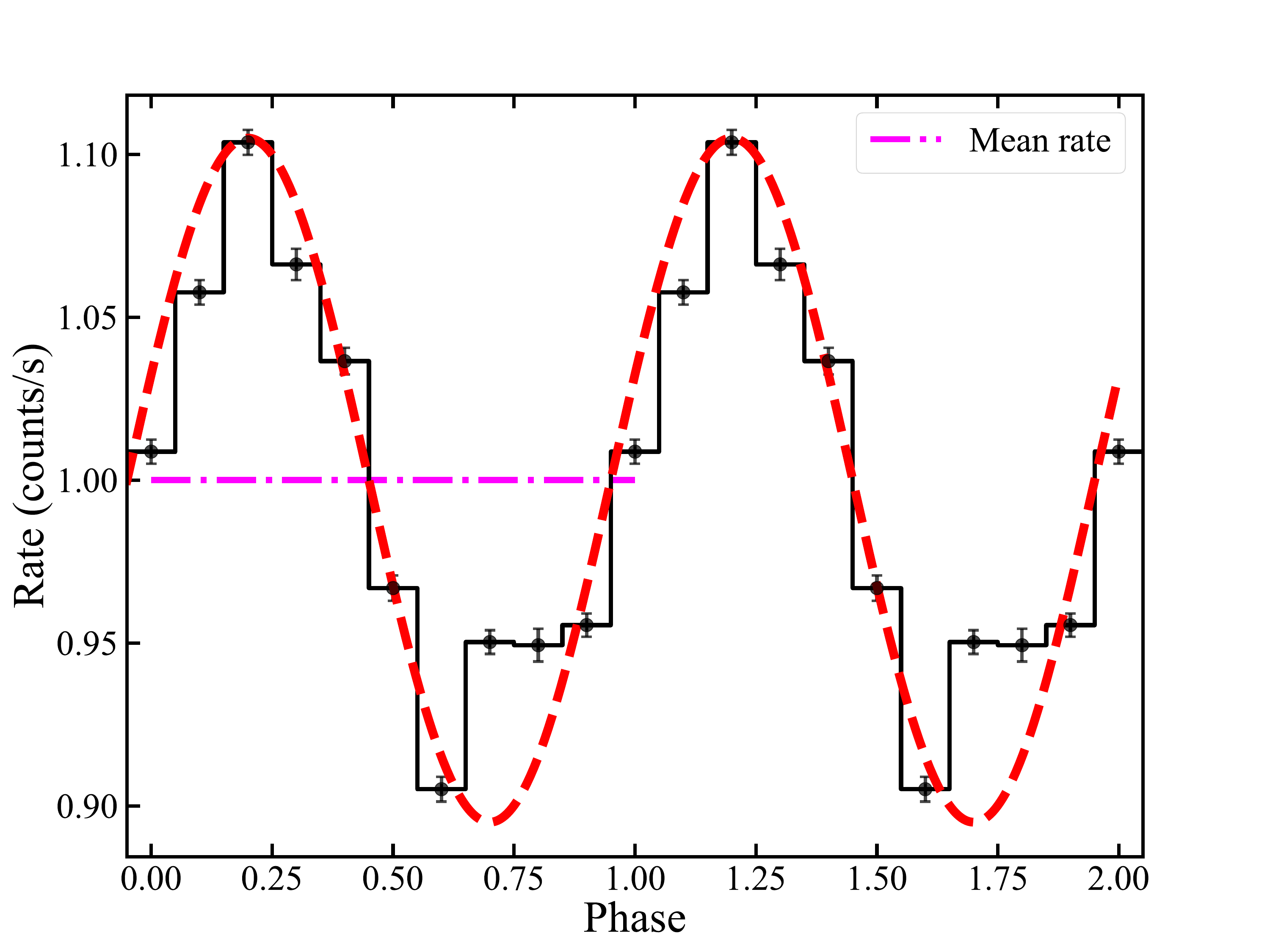}
	\caption{
		Phase-folded light curve with 16000 s with 10 bins in unit phase. The pink dashed-dotted and red dashed lines show the mean count rate and sine-fitted curve, respectively.
	}
	\label{fig:fold}
\end{figure}

\begin{figure}
	\centering
	\includegraphics[width=\linewidth]{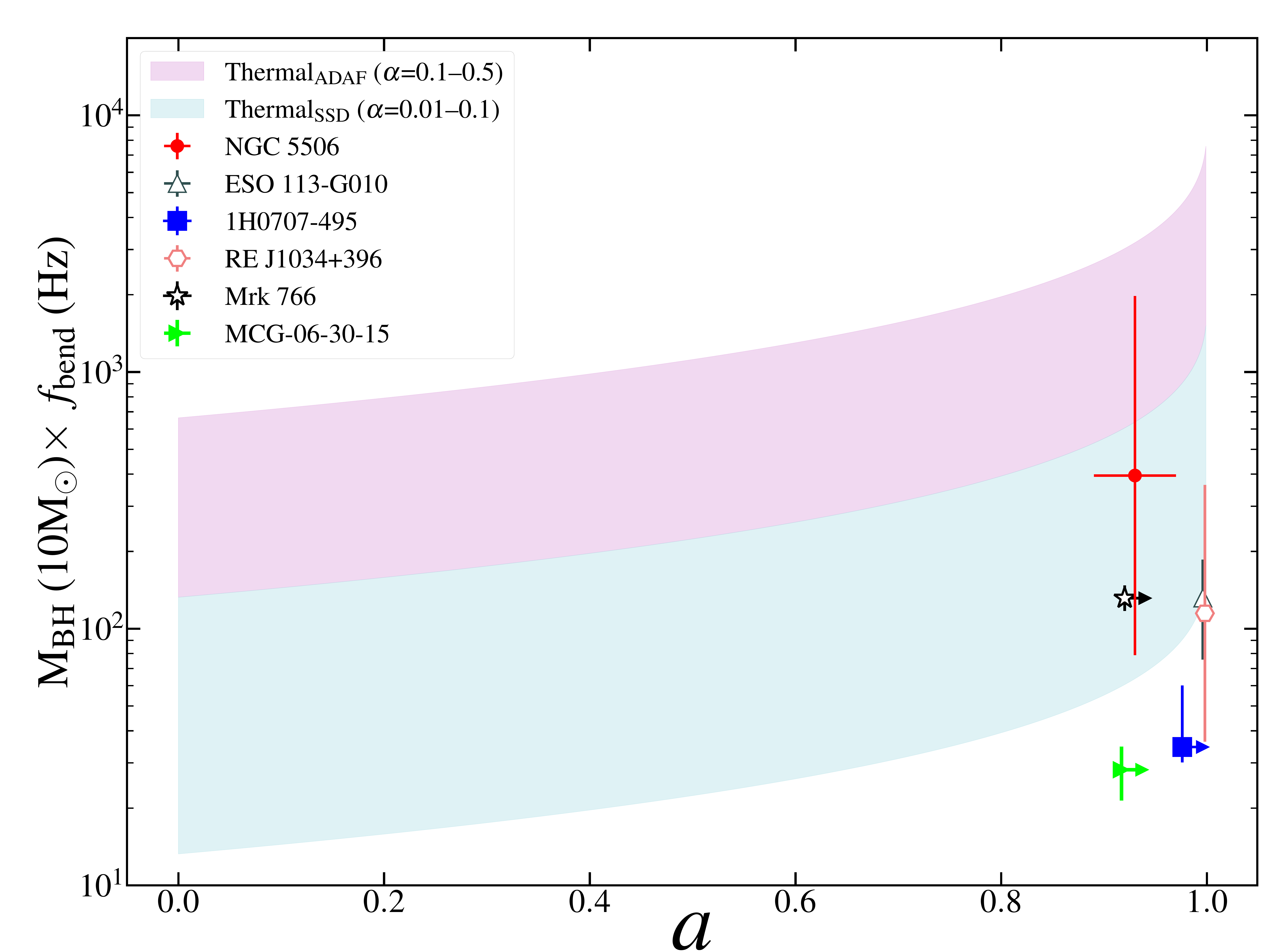}
	\caption{Comparison between thermal and orbital timescales and bending frequency with spin. The black dashed line, sky blue region, and light pink region represent the orbital timescale, thermal$_{\text{SSD}}$, and thermal$_{\text{ADAF}}$, respectively.}
	\label{fig:f_bend}
\end{figure}

\begin{figure*}
	\centering
	\includegraphics[width=\linewidth]{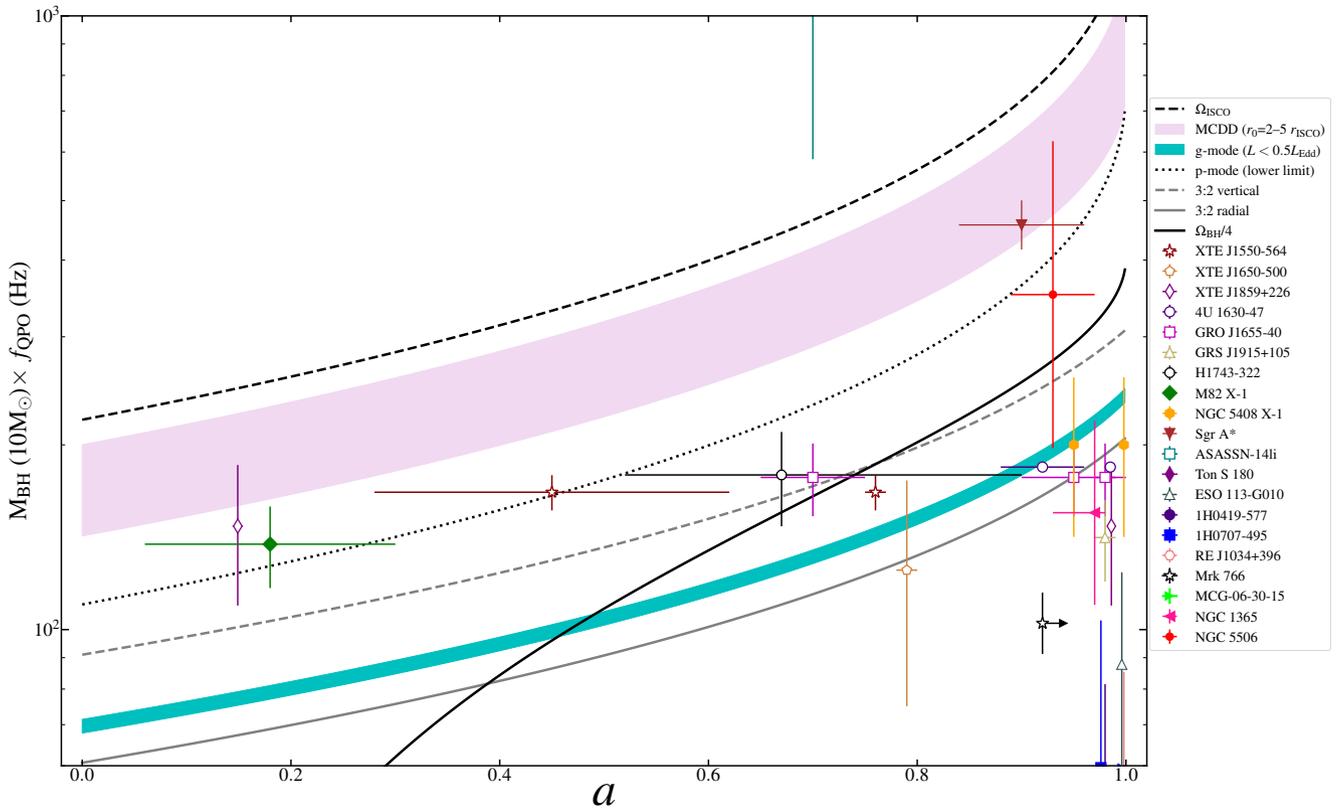}
	\caption{
      Same as Figure \ref{fig:model_fixed}, but with different y-axis scales and the inclusion of NGC 5506 (red point with error bar).
	}
	\label{fig:model_fixed2}
\end{figure*}

\section{A new HFQPO candidate in AGNs}\label{sec:qpo_cand}
In this section, we report a new X-ray HFQPO candidate, NGC 5506, identified from X-ray Multi-Mirror Mission (\textit{XMM-Newton}) observations. This source may exhibit an ADAF configuration similar to that of Sgr A* and NGC 1365, further supporting the HFQPO models.
\subsection{Source information and data reduction}
NGC 5506 is a nearby narrow-line Seyfert~1 galaxy with $z=0.0062$ \citep{2002A&A...391L..21N} at coordinates (J2000) RA=14h13m15s and DEC=-03d12m28s. It is located in edge of the local Virgo supercluster. It is important to note that different methods yield significantly varying results for the central black hole mass of NGC 5506 \citep{2010MNRAS.406.2013G}. We used the result of $10^{7.3\pm{0.7}}M_{\sun}$ in our paper from \citet{2015MNRAS.451.4169G}, and most methods yielded values within this range. The estimate of spin of this source is $0.93\pm{0.04}$ \citep{2018MNRAS.478.1900S}. \citet{2003A&A...402..141B} and \citet{2010MNRAS.406.2013G} reported that the Fe line of this source can be divided into two broad and narrow components. For comprehensive research on the Fe line properties for this source, see \citet{2018MNRAS.478.1900S}. To date, no QPO signals have been reported in this source.

The \textit{XMM-Newton} of the European Space Agency was launched in 1999. It is committed to obtaining high-quality light curves and spectra. For timing analysis, the European Photo Imaging Camera (EPIC)'s PN-CCD \citep{2001A&A...365L..18S} of \textit{XMM-Newton} provides higher-quality light curves than other cameras (MOS; \citealt{2001A&A...365L..27T}) in EPIC. We used the official Scientific Analysis Software (version 19.0.1)\footnote{\url{https://www.cosmos.esa.int/web/xmm-newton/download-and-install-sas}} and standard calibration files\footnote{\url{https://www.cosmos.esa.int/web/xmm-newton/current-calibration-files}} to extract PN photon monitoring data in 0.2--10 (total), 0.2--1 (soft), and 2--10 keV (hard) with 200 s bin$^{-1}$ of each observation of NGC 5506 from the database.\footnote{\url{https://nxsa.esac.esa.int/nxsa-web/\#home}} Then, we strictly followed the standard data reduction thread.\footnote{\url{https://www.cosmos.esa.int/web/xmm-newton/sas-threads}} The background count-rate threshold was set as RATE \textless=0.4. The source range was set to 40$\arcsec$ and 50$\arcsec$ with a circular CCD chip  for small and large window modes, respectively. In small window mode, the background region was far away from source with the same shape as the source. The background region was an annulus of large window mode, ranging from 50$\arcsec$ to 100$\arcsec$. No observations had pile-up effects.

\subsection{Periodic analysis}
In ObsID:0554170201 (starting on 2008-07-27 with a good time interval of $\sim$90 ks), we detected an obvious peak in the power spectrum. The results of total band are shown in Figure \ref{fig:wwz_result}. We combined the results of the Lomb--Scargle periodogram (LSP; \citealt{1976Ap&SS..39..447L,1982ApJ...263..835S}) and the weighted wavelet Z-transform (WWZ; \citealt{1996AJ....112.1709F}) into one figure. As shown in panel (C), there is a clear peak (red dashed line) at $16024 \pm 1345$ s in the LSP, and the peak position in the time-averaged WWZ power spectrum (black solid line) is almost the same as LSP. The uncertainty of peak locations is obtained by the full width at half maxima of the Gaussian function to fit the peak. In panel (A), the red arrows indicate the peak positions of the periodic oscillation in light curve. In order to evaluate the significance of the peak, we used Monte Carlo simulations\footnote{\url{https://github.com/samconnolly/DELightcurveSimulation}} \citep{2013MNRAS.433..907E} to generate 10$^{6}$ artificial light curves. Since the power spectrum of NGC 5506 exhibits a low-frequency bending \citep{2012A&A...544A..80G}, the red noise background should be modeled as a bending powerlaw (BPL): $P(f)=Af^{-1}[1+(f/f_{\text{bend}})^{\alpha-1}]^{-1}+C$. The best fit values for $A$, $\alpha$, $f_{\text{bend}}$ and $C$ are $(3.3\pm{1.2})\times10^{-3}$, $3.77\pm{0.7}$, $(1.98\pm{0.6})\times10^{-4}$ and $0.27\pm{0.026}$, respectively. The peak of LSP is at the 3.6$\sigma$ significance level (green dash-dotted line). The light curves of AGNs can be modeled as a damped random walk (DRW) process \citep{2009ApJ...698..895K,2010ApJ...721.1014M,2013ApJ...765..106Z,2019PASP..131f3001M,2024ApJ...967L..18Z}. Thus, this model can be used to describe background red noise and estimate the significance of QPO signals \citep{2024arXiv241106366R}. We simulated $10^{6}$ light curves using the \texttt{celerite} tool\footnote{\url{https://celerite.readthedocs.io/en/stable/}} developed by  \citet{2017AJ....154..220F} and obtained a peak significance of 3.7$\sigma$, shown as blue dashed line in panel (C) of Figure \ref{fig:wwz_result}. Both methods gave almost identical results. The 2D time-frequency power map (panel (B) of Figure \ref{fig:wwz_result}) shows that the QPO signal is stable and persistent in the total observation. In this observation, the soft band has no periodic signal. The light curve and periodic signal in the hard band are almost the same as the total. The results of the hard and soft band are shown in Figure \ref{fig:wwz2}. For the soft band, since the DRW model does not fit the light curve well, we only give the significance level of BPL.

Finally, we derived a QPO signal with $\sim$16000 s with the above methods. We used this QPO value to fold the light curve using the \texttt{efold} tool\footnote{\url{https://heasarc.gsfc.nasa.gov/docs/xanadu/xronos/examples/efold.html}} with 10 bins in unit phase. In Figure \ref{fig:fold}, the phase-folded light curve shows a periodic shape in two phases. This once again demonstrates the reliability of the QPO. The number of different bins in unit phase may slightly affect the results, but will not alter the periodic shape.

\subsection{Model constraint}
Given the large uncertainty in the black hole mass of NGC 5506, further evidence is required to constrain its mass estimate. \citet{2012A&A...544A..80G} reported that some AGNs exhibit low-frequency bending in their X-ray power spectra, with six sources overlapping with our HFQPO catalog (including NGC 5506). The bending frequency is thought to originate from a characteristic timescale of some physical processes around the ISCO \citep{2021Sci...373..789B}, such as thermal time (the time scale to restore thermal equilibrium). The thermal timescale at $r_{\text{ISCO}}$ with $M$ is as follows \citep{2024ApJ...967L..18Z}:
\begin{equation}
	t_{\text{thermal}}= 1.45\left(\frac{\alpha}{0.01}\right)^{-1}\left(\frac{r_{\text{ISCO}}}{200r_{\text{g}}}\right)^{3/2}M_{\text{BH}}(M_{\sun}) \ \text{s},
\end{equation}
where $\alpha$ is the viscosity parameter of accretion flow. In ADAFs, significant advection results in an $\alpha$ value between 0.1 and 0.5 \citep{1998tbha.conf..148N,2012A&A...545A.115K,2013ApJS..207...17L}. By contrast, the SSD typically exhibits a smaller $\alpha$ value, ranging from 0.01 to 0.1 \citep{2010ApJ...713...52D,2013ApJ...772..102H,2019NewA...70....7M}. Similar to the HFQPO model, the effect of black hole spin can be incorporated into this model. As shown in Figure \ref{fig:f_bend}, there is reasonable agreement between certain HFQPO sources with bending frequencies and the theoretical models, with 1H0707-495 and MCG-06-30-15 favoring lower viscosity parameter values. Most HFQPO sources support an SSD configuration, which is consistent with the conclusions of Section \ref{sec:obs_the}. NGC 5506 is located near the boundary between the ADAF and SSD regimes. Its Eddington ratio is $0.05^{+0.21}_{-0.04}$ \citep{2024A&A...686A..46E}, significantly lower than the average value of $\sim 0.43$ observed in other AGN-QPO sources. Notably, NGC 5506 also exhibits a sub-parsec bent radio jet \citep{2000ApJ...537..152K,2024A&A...686A..46E}. These characteristics suggest that NGC 5506 likely harbors an ADAF structure, with the black hole mass being close to the upper limit of the measurement uncertainty. Based on this, we provide a further constrained estimate of the black hole mass and Eddington ratio as $10^{7.75\pm{0.25}} M_{\sun}$ and $0.02^{+0.01}_{-0.01}$, respectively.

As shown in Figure \ref{fig:model_fixed2}, NGC 5506 occupies the same parameter space as ULXs and XRBs, supporting the existing HFQPO models. This source also exhibits an ADAF configuration similar to that of Sgr A* and NGC 1365.

\section{Discussion} \label{sec:dis}
As SIAP faces challenges, it is crucial to incorporate more observations for comparisons with the models. We compiled an updated X-ray HFQPO catalog from recent studies that include sources with various mass scales. Spin is a key parameter in accretion disk and black hole physics \citep{1972ApJ...178..347B}, and incorporating black hole spin into theoretical models provides more accurate results regarding SIAP than models without this parameter. Our analysis shows that ULXs support the models we invoked, similar to XRBs. There is no significant difference between sources with or without harmonics, though it is worth noting that most harmonics are non-simultaneous. Many AGNs still do not support these models, but we found that Sgr A* and NGC 1365 do support them. Both sources have a high probability of containing an ADAF configuration with a low accretion rate. Additionally, we detected a possible HFQPO in NGC 5506 at $\sim$16000 s with significance of 3.6$\sigma$ by BPL and 3.7$\sigma$ by DRW. This source also exhibits a low accretion rate pattern with a jet structure, consistent with an internal ADAF configuration. The detection of HFQPO in NGC 5506 exclusively in the hard X-ray band provides additional evidence for the presence of an ADAF. Like Sgr A* and NGC 1365, NGC 5506 supports the existing HFQPO models, which is equivalent to supporting SIAP. This finding may suggest that the accretion state in SMBHs that makes the HFQPO model effective differs from that in SBHs and IMBHs and is consistent with the results of numerical simulations \citep{2012ApJ...746L..10D,2013ApJ...774L..22S}. In addition, the influence of accretion states on QPOs exhibits consistent patterns between EUV and X-ray signals.

Previous studies have focused on searching for QPO signals in narrow-line Seyfert 1 galaxies, as these AGNs tend to have long archival X-ray obervations that enable QPO discovery. However, our results suggest that QPOs are more common in AGNs than previously thought and accretion states are the key factors influencing QPO. Therefore, the selection effects could be a significant factor contributing to the previous discrepancy between the observation and theory. Many studies have focused on unifying the physical structure of SBH and SMBH accretion systems (e.g., \citealt{2006A&A...456..439K, 2006Natur.444..730M, 2010LNP...794.....B, 2015ApJ...798L...5Z, 2015SciA....1E0686S, 2021Sci...373..789B, 2024ApJ...967L..18Z}). However, it is evident that there are some notable differences between these two types of accretion systems. For example, spectral state transitions generally occur only in XRBs \citep{2006ARA&A..44...49R}, transient jet structures in XRBs (related to spectral transitions), the soft X-ray excess is common in AGNs but rare in XRBs \citep{2021MNRAS.500.2475J}, and XRBs have a higher disk electron density than AGNs \citep{2023ApJ...951..145L}. Since most of the current models are toy models, a global general-relativistic magnetohydrodynamic (GRMHD) HFQPO model that accounts for these differences between black hole accretion systems at various scales and differences between different AGN subclasses may be key to reconciling many of the AGN rejection models. Given that the central black hole in XRBs is of stellar mass, the accretion state can change rapidly, making it challenging to fully understand the underlying physics. Otherwise, as \citet{2021ApJ...906...92S} suggest, those HFQPOs that do not conform to the models may originate from other physical processes, such as the EMRI model. Our assessment of the available evidence supports SIAP as the fundamental framework for describing the accretion process of black holes. Additionally, long-term, multi-band monitoring is essential for further progress.

\section{Conclusion} \label{sec:con}
We compiled an updated X-ray HFQPO catalog encompassing SBH, IMBH, and SMBH accretion systems and revisited the observational and theoretical confrontations. Our analysis demonstrates that ULXs support the models we invoked, similar to XRBs. We identified two SMBH sources with possible ADAF configurations that can support the existing HFQPO models, even though many AGNs still do not support them. This result is consistent with numerical simulations. Additionally, we detected a possible HFQPO in NGC 5506, which supports the models and exhibits a potential ADAF configuration. Given that the SIAP framework always holds true, further GRMHD HFQPO models that can incorporate both the similarities and differences in the accretion structures of SBHs and SMBHs are crucial for reconciling AGN rejection models. An alternative explanation is
that HFQPOs that reject the models originate from other physical processes. Thus, the paradigm of accretion scale invariance remains consistent from SBHs to SMBHs. More observational evidence is needed to enrich this framework in the future.

\section*{Acknowledgements}
We thank the anonymous referee for providing constructive
comments and suggestions. This work was partly supported by the National Science Foundation of China (12263007 and 12233006) and the High-level Talent Support Program of Yunnan Province. This work was based on observations conducted by XMM-Newton, an ESA science mission with instruments, and contributions directly funded by ESA Member States and the USA (NASA).

\section*{Data Availability}
The data that support the findings of this study are available from the corresponding author upon reasonable request.
\bibliographystyle{mnras}
\bibliography{myBiblio}
\end{document}